\documentclass{ws-procs9x6}
\newcommand{\bce}{\begin{center}} \newcommand{\ece}{\end{center}}
\newcommand{\beq}{\begin{equation}} \newcommand{\eeq}{\end{equation}}
\newcommand{\beqy}{\begin{eqnarray}}
\newcommand{\eeqy}{\end{eqnarray}}
\begin{document}

\title{Distinguishing Bare Quark Stars from Neutron Stars}

\author{P. JAIKUMAR$^1$\footnote{\uppercase{W}ork presented
at the 26th Annual 
\uppercase{M}ontreal-\uppercase{R}ochester-\uppercase{S}yracuse-\uppercase{T}oronto
Conference (\uppercase{MRST} 2004) on High Energy Physics: \uppercase{F}rom Quarks to 
Cosmology, \uppercase{M}ay 12 - 14 (2004) \uppercase{C}oncordia \uppercase{U}niversity, 
\uppercase{M}ontr\'{e}al, \uppercase{QC}, \uppercase{C}anada}~\,\footnote{Electronic address: jaikumar@hep.physics.mcgill.ca} 
~, C. GALE$^1$, D. PAGE$^2$, 
        AND M. PRAKASH$^3$}

\address{$^1$Physics Department, McGill University, Montr\'{e}al,
         Qu\'{e}bec H3A 2T8, Canada.\\
 	 $^2$Instituto di Astronomia, UNAM, Mexico D.F. 04510, Mexico.
         \\              
         $^3$Department of Physics \& Astronomy, SUNY at Stony Brook,
        Stony Brook, NY 11794, USA.}

\maketitle

\abstracts{Observations to date cannot distinguish neutron stars from
self-bound bare quark stars on the basis of their gross physical 
properties such as their masses and radii alone.
However, their surface luminosity and spectral characteristics
can be significantly different. Unlike a normal neutron star,  
a bare quark star can emit photons from its surface at super-Eddington
luminosities for an extended period of time. We present a calculation  
of the photon bremsstrahlung rate from the bare quark star's surface,
and indicate 
improvements that are required for a complete
characterization of the spectrum. The observation of this distinctive
photon spectrum would constitute an unmistakable signature of a
strange quark star and shed light on color superconductivity at
stellar densities.}

\section{Introduction}

A normal neutron star (NS), composed mainly of neutron matter, is
bound by gravitational forces. On the other hand, a strange quark
matter (SQM) star, consisting of three-flavor ($u,d,s$) quark matter
up to its surface, can be self-bound due to strong interactions
alone. This makes the pressure at the star's surface vanish at a
supra-nuclear baryon density. In the context of the MIT bag model with
first order corrections due to gluon exchange, the baryon density
$n_B$ at which the pressure vanishes is~\cite{PBP90}
\beq
\label{nzero}
n_B(P=0)=\biggl(\frac{4B}{3\pi^{2/3}}\biggr)^{3/4}
\biggl(1-\frac{2\alpha_c}{\pi}\biggr)^{1/4},
\eeq
where $B$ is the bag constant and $\alpha_c=g_c^2/(4\pi)$ is the
quark-gluon coupling constant. For typical values of $B$ and
$\alpha_c$~\cite{Farhi84}, the baryon density at vanishing pressure is about
2 to 3$n_0$, where $n_0=0.16$~fm$^{-3}$ is the saturation density of
nuclear matter. 

\vskip 0.1cm

\noindent Because of differences in the 
cause for binding, the mass versus radius relations of NS and SQM
stars differ significantly, although in the range of masses
($1<M/M_{\odot}<2$) observed to date \cite{LP04}, the calculated radii
are similar ($R\sim 10$ to 15 km). This makes it difficult to
distinguish these two classes on the basis of their gross physical
properties alone. However, the light curves of NS and SQM stars,
determined by surface photon emission, can be very different. 
At the surface of a SQM star, the density drops
abruptly to zero within a distance of the order of a few fm, and
charge neutrality 
requires an electron concentration
$n_e/n_B$ of $10^{-4}$ to $10^{-3}$. These electrons at the surface
are bound by electrostatic interaction to quark matter, with radial
electric fields whose magnitude ($\sim 5\times 10^{17}~{\rm
V~cm}^{-1}$) can exceed the critical value 
for electron-positron pair production from the QED vacuum.

\vskip 0.1cm

\noindent Recently, it has been pointed out that thermal emission from the bare
surface of a strange quark star, due to 
$e^+e^-$ pair production and subsequent annihilation to photons, can
produce luminosities well above the Eddington limit ($\sim
10^{38}~{\rm erg~sec}^{-1}$) for extended periods of time, from about
a day to decades~\cite{Page02}. The spectrum of 
photons is
different from that of a normal cooling neutron star ($30
<E/{\rm keV} < 500$ instead of $0.1 < E/{\rm keV} <2.5$). Identifying
this distinctive 
spectrum is well within the capabilities of
the INTEGRAL satellite~\cite{Integral} launched toward the end of
2002.

\section{Bremsstrahlung photons from the surface of a bare quark star} 

In addition to $e^+e^-$ pair production at the surface, photons are
emitted from the bremsstrahlung process $e^-e^- \rightarrow
e^-e^-\gamma$ in the electron layer, termed the ``electrosphere''. We
report a calculation of the corresponding emissivity in the soft
photon limit and compare its magnitude to those from pair-production and
blackbody emission. In addition, we characterize the spectrum by its mean
energy and quantify deviations from a blackbody spectrum.
Our results apply to the case 
in which there is a charge neutralizing layer of electrons at the
surface, including 
superconducting phases of quark matter~\cite{AR02}.

\vskip 0.1cm

\noindent As the surface temperature falls below $10^{10}$ to $10^{11}$K
($\sim (1-10)$ MeV) within tens of seconds of a compact star's birth, 
electrons quickly settle into a degenerate Fermi sea, and their Fermi
momentum far exceeds their rest mass, making it a relativistic
degenerate system. In the plane-parallel approximation for the layer
of the electropshere, the electron chemical potential as a function of
distance $z$ from the quark surface is given by~\cite{NKG95}
\beq
\mu_e(z)=\frac{\mu_e(0)}{1+z/H};\quad H=\frac{\hbar~c}{\mu_e(0)}
\sqrt{\frac{3\pi}{2\alpha}}=501.3
\biggl(\frac{10~{\rm MeV}}{\mu_e(0)}\biggr)~{\rm fm}\,,\label{eprofile}
\eeq
where $\alpha=e^2/4\pi$ is the fine structure constant. For a typical
$\mu_e(0)=(10-20)$ MeV, the electrosphere is about $10^3$~fm thick or
more.  The variation of $\mu_e$ across the electrosphere is shown in
Fig.~1. The scattering of electrons in the electrosphere 
gives rise to bremsstrahlung photons.
\begin{figure}[!ht]
\bce
\includegraphics[width=5.0cm,height=6.0cm,angle=270]{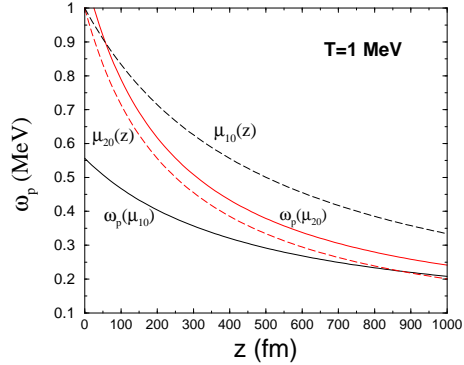}
\ece
\caption{Profiles of the electron chemical potential $\mu_e$
(Eq.~(\ref{eprofile})) and the plasma frequency $\omega_p$ in the
degenerate limit (Eq.~(\ref{omegap})) versus distance $z$ from the
surface of the strange quark matter star. $\mu_{10}~(\mu_{20})$ is the
electron chemical potential in units of 10 (20) MeV. }
\label{figps1}
\end{figure} 
The emitted photons have 
to travel through the electron plasma, therefore, 
their in-medium dispersion relation can be taken as
$\omega=(\omega_p^2+k^2)^{1/2}$ (in $c=1$ units), where the plasma frequency
$\omega_p$ serves as a low-energy cut-off. For degenerate electrons
($T/\mu_e \ll 1$), the plasma frequency is determined
from~\cite{SSP03}
\beq
\label{omegap}
\omega_p^2\cong\frac{4\alpha}{3\pi}
\mu_e^2\biggl(1+\frac{\pi^2T^2}{3\mu_e^2}\biggr) \,.
\eeq
The optical depth of the electrosphere at any distance is
determined by $\omega_p$,  
which decreases with
increasing $z$ (see Fig.~1). The emissivity $Q$ is thus a function of
$\omega_p(z)$. The total luminosity can be expressed as
\beq
L = 4\pi R^2\int_{z=0}^{z=z_0}Q(\omega_p(z))\,dz \,,
\eeq
where $R$ is the star's radius and $z_0$ is the thickness of the
electrosphere. As quark matter is admixed with electrons in the
innermost region of the electrosphere, the exchanged photon is
screened. We take this into account 
by modifying
the photon propagator appropriately. Our results demonstrate that
electric screening and magnetic damping effects play only a small role
for the emission of low energy photons.

\section{Calculation of Photon Emissivity}

The photon emissivity from the bremsstrahlung process is
\beqy
\label{emiss1} 
 Q = \frac{2\pi}{s\hbar}&&\biggl[\prod_{i=1}^{4}\int\frac{d^3p_i}{(2\pi)^32\omega_{p_i}}\biggr]\int\frac{d^3k}{(2\pi)^32\omega_k}\omega_k\sum_{spin}|M|^2~S_{LPM}(k)\\  \nonumber
&\times&n_F(\omega_{p_1})n_F(\omega_{p_2})
  \tilde{n}_F(\omega_{p_3})\tilde{n}_F(\omega_{p_4})(2\pi)^3\delta^3
({\bf P_f-P_i})\delta(E_f-E_i)\quad. 
\eeqy
The subscripts $i=1$ to 4 refer to electrons, $(\omega_k,{\bf k})$ is
the 4-momentum of the emitted photon.\footnote{In the low-energy
limit, the emission from the exchange diagrams simply doubles the
overall result for the emissivity from the direct diagrams (no
interference). Since there is a symmetry factor of $s=2$ in the
denominator of the emissivity expression Eq.~(\ref{emiss1}), we may
ignore the identity of the particles altogether and evaluate the
emissivity from the direct diagrams alone.}. The phase space for
electrons is convoluted with the appropriate Fermi distribution
functions $n_F(\omega_{p_i})=1/({\rm e}^{\omega_{p_i}/T}+1)$ (in
$k_B=1$ units) and ${\tilde n_F}=1-n_F$, respectively. Pauli blocking
for degenerate fermions restricts the intermediate state to be almost
on shell, so that the contribution of low-energy photons dominates the
emissivity. $S_{LPM}$ is a suppression factor due to the
Landau-Pomeranchuk-Migdal (LPM) effect of multiple scattering during
emission, which we estimate numerically. For this section, we set
$S_{LPM}=1$ to obtain useful analytical expressions for the
emissivity. We work in Low's approximation which is accurate for
low-energy photons. In this limit, the full scattering amplitude $M$
for bremsstrahlung factorizes into an elastic part (M$\o$ller
scattering) and a part which gives the classical intensity 
divided by the energy $\omega_k$.

\vskip 0.1cm

\noindent In the soft photon limit, the analytic expression for the emissivity
$Q$ in two relevant regimes of temperature~\footnote{When $T\sim
m_e^2/\mu_e$, recoil effects become important and the matrix element
is more complicated.} reads
\beqy 
\label{finale} 
Q&=&\frac{64\alpha^3m_e\mu_e^3{F}(\omega_p)}{15\hbar(2\pi)^6}{\rm
ln}\biggl(\frac{\mu_e}{m_e}\biggr){
I}(T,\mu_e) \label{efinal}\\ \nonumber 
{F}(\omega_p) &=&
\biggl\{1+\frac{1}{2}\frac{\omega_p^2}{\omega_p^2+m_e^2}-\frac{3\omega_p}{2m_e}{\rm tan}^{-1}\biggl(\frac{m_e}{\omega_p}\biggr)\biggr\} 
\\ \nonumber
{I}(T,\mu_e) & = & 
8\pi^2\times\left\{\begin{array}{c}\frac{4T}{\mu_e^3}\biggl({\rm ln}~2-
\frac{1}{2}\biggr);\quad
\frac{m_e^2}{2\mu_e}\leq T \ll 
\mu_e \\ \frac{2}{\mu_e^2}{\rm e}^{-m_e^2/2\mu_e T};\quad
T\ll\frac{m_e^2}{2\mu_e}\end{array}\right\} \,.
\eeqy 
%
Above, terms of order $\left(m_e^2/m_{Debye}^2\right)$ 
arising from
electric screening effects have been ignored 
because they are small. The factor  
${\rm ln}\left(\frac{\mu_e}{m_e}\right)$ comes from forward scattering
mediated by unscreened magnetic gluons. Thus, screening effects are
sub-dominant in the low-energy limit. In order to test the accuracy of
the analytic result, we performed a 
numerical integration of
Eq.~(\ref{emiss1}), and a comparison of analytical and numerical results, 
along with the pair-production rate~\cite{Usov01} is shown in Fig.~2.

\begin{figure}[!ht]
\bce
\includegraphics[width=5.0cm,angle=270]{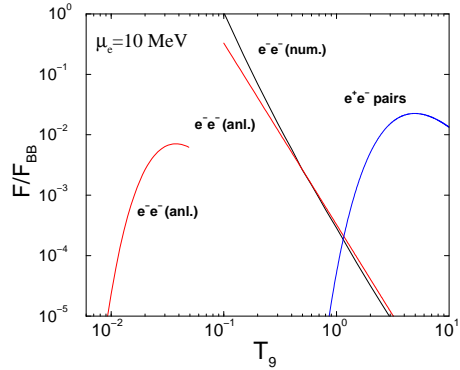}
\ece
\caption{Photon flux from the bremsstrahlung and pair annihilation
processes scaled to the blackbody limit as a function of temperature
in units of $10^9$K ($T_9$). The dashed curve (num.) shows results of
numerical integrations of Eq.~(\ref{emiss1}) which is to be compared
with the results of the analytical expressions (anl.) from
Eq.~(\ref{finale}) for low and high temperatures. The (uniform)
electron chemical potential is $\mu_e=10$ MeV with $z_0=100$ fm.}
\label{figps2}
\end{figure}

\noindent The flux of bremsstrahlung radiation 
dominates over that
from pair-creation for electron plasmas below a temperature of $\sim
0.1$ MeV. Its dependence on the third power of the
electron chemical potential (or linear in electron density) 
distinguishes it from other processes.

\vskip 0.1cm

\noindent The spectrum of low energy bremsstrahlung photons is given by~\cite{JPPG04}
\beq
\label{spectrum}
\omega_k\biggl(\frac{d\sigma}{d\omega_k}\biggr) \propto 
\biggl(1-\frac{\omega_p^2}{\omega_k^2}\biggr)^{3/2} ~~
{\rm ln}\biggl(\frac{\mu_e^2}{m_e\omega_k}\biggr)\quad . 
\eeq
In the quasiclassical approximation, support for this spectrum exists
only between $\omega_{min} = \omega_p$ and $\omega_{max} = (\omega_p^2
+ m_e^2)^{1/2}$. One must go beyond this in order to obtain the
spectrum for $\omega > \omega_{max}$.  Equation (\ref{spectrum}) 
shows that the presence of the electron plasma produces a harder
spectrum than in vacuum, whereas the total photon flux is
smaller. Although the contribution of high energy photons cannot be
reliably estimated within our approximations, it is expected to fall
off steeply due to the large electron degeneracy.
\begin{figure}[!ht]
\leavevmode \bce
\includegraphics[width=4.2cm,angle=270]{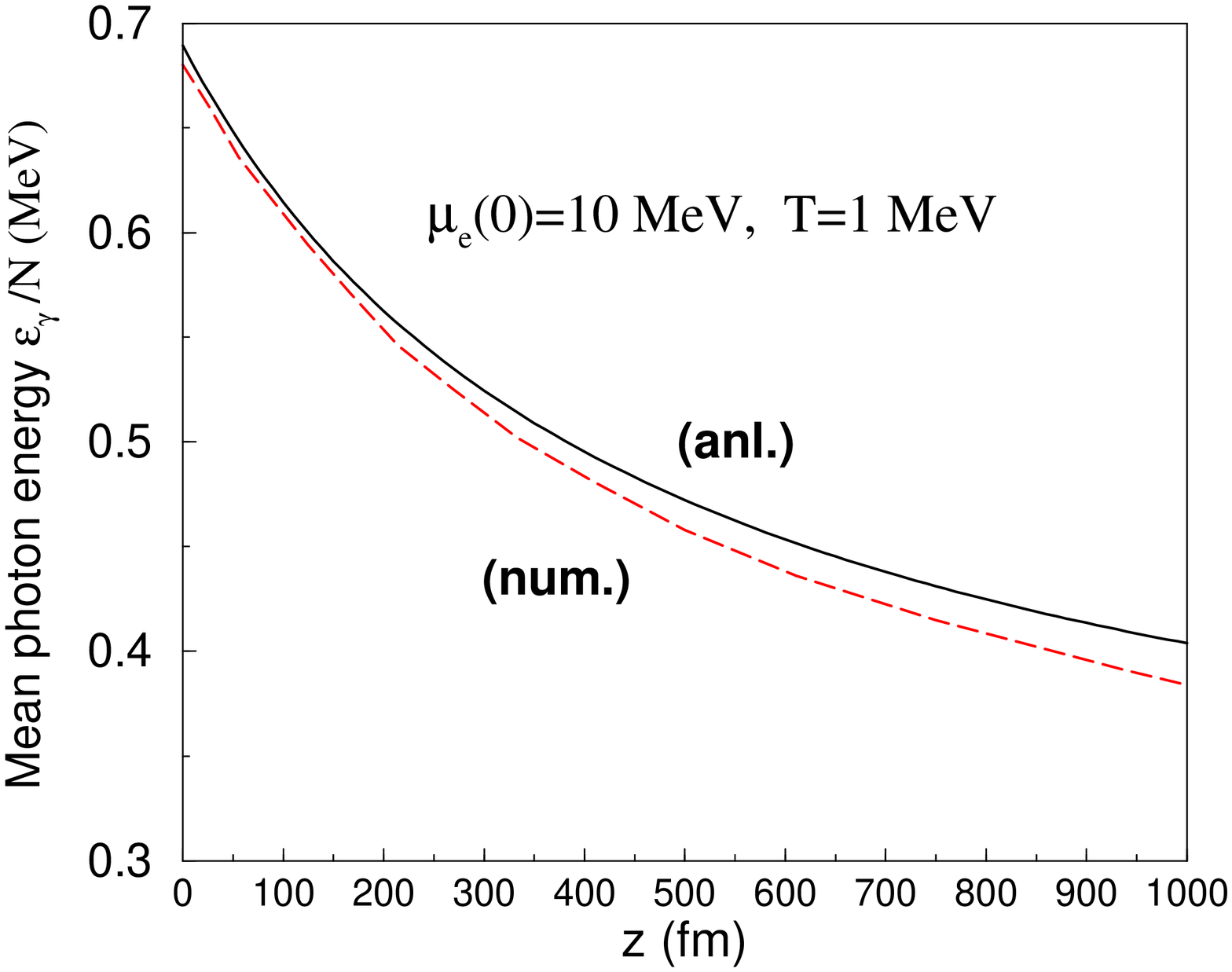}
\includegraphics[width=4.2cm,angle=270]{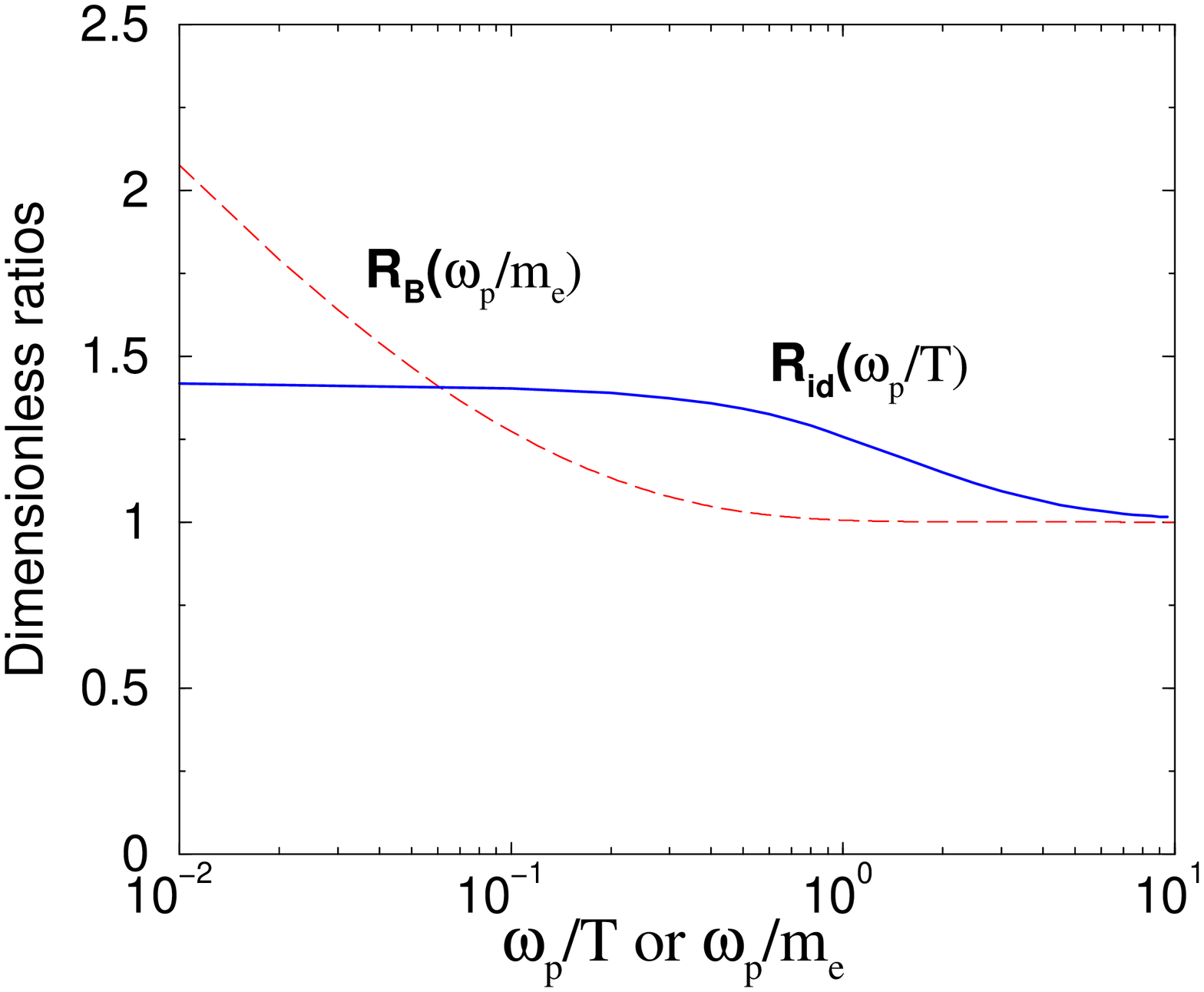}
\caption{(left panel): Mean photon energy versus $z$, calculated
analytically (anl) and numerically (num). (right panel): A comparison
of the ratio $R$ in Eq.~(\ref{rat}) for bremsstrahlung photons and
$R_{id}$ for thermal photons with in-medium photon mass $\omega_p$. }
\ece
\label{figR}
\end{figure}

\noindent The mean photon energy from bremsstrahlung is given by (see
Fig.~\ref{figR})
\beq
\langle \omega \rangle = {Q}/{\Gamma} \approx 0.5 {~\rm MeV}\,,
\eeq
where $\Gamma$ is the number of bremsstrahlung reactions per unit
volume per unit time (the rate). Deviations from a blackbody spectrum are
quantified by the dimensionless ratio 
\beq 
\label{rat}
R_B = {\langle \omega^2 \rangle} / {\langle \omega \rangle^2} \,,
\eeq  
where $\langle \omega^2 \rangle = {Q^2}/{\Gamma}$. Figure~\ref{figR}
contrasts this ratio for the bremsstrahlung spectrum against ideal gas
behavior $R_{id}$ . The non-thermal feature of the bremsstrahlung
photons is due to their large mean free paths 
in the electrosphere, so that they do not scatter often enough to
thermalize. In light of the limited frequency support within our
low-energy approximations, a more rigorous quantum calculation is
called for to more accurately characterize the spectrum, which depends
primarily on the ratio $\omega_p/m_e$. 

\vskip 0.1cm

\noindent Multiple scattering of electrons within the formation time of the
radiated photon can lead to Landau-Pomeranchuk-Migdal (LPM)
suppression of bremsstrahlung. A numerical integration of
Eq.~(\ref{emiss1}) including the suppression factor $S_{LPM}(k)$
reduces the simple analytical estimate of
Eq.~(\ref{finale}). Specifically, for $\mu_{10}=1$, the reduction
factor (relative to the emissivity without the LPM effect) is $\sim
60$, while for $\mu_{10}=2$, the reduction factor is $\sim 350$.
The mean photon energy and spectral shape do not change much upon the
inclusion of the LPM suppression 
because both the rate and the energy loss are affected similarly, and
nearly cancel in the ratio.
\begin{figure}[!ht]
\bce
\includegraphics[width=5.0cm,height=9.0cm,angle=270]{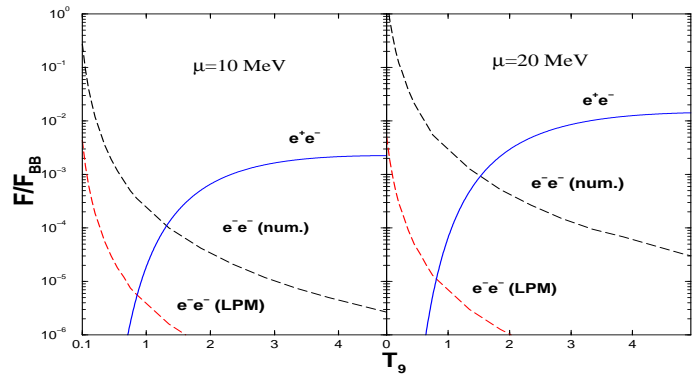}
\ece
\caption{Photon flux from the bremsstrahlung and pair
annihilation processes scaled to the blackbody value as a
function of temperature in units of $10^9$K ($T_9$).  In the case of
the bremsstrahlung process, the upper (lower) curve shows results
without (with) the effects of the LPM effect. }
\label{figlpm}
\end{figure}

\section{Importance for cooling of bare quark stars}

The emitted photons rob  
energy from the low 
temperature electron plasma and cause the surface of the quark star to cool.  
The bremsstrahlung rates calculated 
in this work, together with the rates of pair annihilation processes
calculated earlier, can be used as input for calculations of the
thermal evolution of strange quark matter stars.
Improved estimates of the bremsstrahlung emissivity applicable in the
temperature range
excluded by Eq.~(\ref{finale}) require 
a proper account of photon absorption processes,
as bremsstrahlung emission becomes comparable
to blackbody emission. However, for 
$T/^{o}{\rm
K}\leq 10^{9}$, bremsstrahlung is the dominant photon emission
process, 
as the luminosity from pair annihilation is
vanishingly small. The former also dominates equilibrium and
non-thermal bremsstrahlung radiation from quark-quark collisions in
the uppermost layer of quark matter~\cite{Cheng03}. 
As the temperature falls below $10^7$ K, the rise in the
bremsstrahlung rate is halted by the exponential factor in
Eq.~(\ref{finale}) which causes the emissivity to turn over and rapidly 
decrease. 
For $T<10^9$K, the bremsstrahlung process dominates photon emission
even in the presence of LPM effects, suggesting that it will be
important in baseline calculations of the surface luminosity (or
temperature) versus age of bare quark stars that can help distinguish
them from neutron stars.

\vskip 0.1cm

\section*{Acknowledgments} 
P.J. and C.G. are supported in part by the Natural Sciences and Engineering
Research Council of Canada and in part by the Fonds Nature et Technologies of
Quebec. The research of M.P. was supported by the U.S. Department of
Energy grant DOE/DE-FG02-87ER-40317 and the NSF grant INT-9802680. The
work of D.P. is partially supported by grants from UNAM-DGAPA
(PAPIIT-IN112502) and Conacyt (36632-E). \rm


\begin{thebibliography}{0}

\bibitem{PBP90}
Manju Prakash, E. Baron and M. Prakash, Phys. Lett. {\bf B243}, 75 (1990).
\bibitem{Farhi84}
E. Farhi and R. L. Jaffe, Phys. Rev. {\bf D30}, 2379 (1984).
\bibitem{LP04}
J. M. Lattimer and M. Prakash, Science {\bf 304}, 536 (2004). 
\bibitem{Page02}
D. Page and V. V. Usov, Phys. Rev. Lett. {\bf 89}, 131101 (2002). 
\bibitem{Integral}
See, e.g., J. Knodlseder, astro-ph/0207527.
\bibitem{AR02}
M. Alford and K. Rajagopal, 
J. High Energy Physics, {\bf 0206} 031 (2002).
\bibitem{NKG95}
C. Kettner, F. Weber, M. K. Weigel and N. K. Glendenning,
Phys. Rev. {\bf D51}, 1440 (1995).
\bibitem{SSP03}
S. Ratkovi\'c, S. I. Dutta and M. Prakash, Phys. Rev. {\bf D67},
123002 (2003).
\bibitem{JPPG04}
P. Jaikumar, C. Gale, D. Page and M. Prakash, Phys. Rev. {\bf D}, in press.
\bibitem{Usov01}
V.V. Usov, Astrophys.J. {\bf 550}, L179 (2001).
\bibitem{Cheng03}
K. S. Cheng and T. Harko, Astrophys.J. {\bf 596}, 451 (2003). 

\end{thebibliography}
\end{document}